\documentclass{article}
\usepackage[top=1.5in, bottom=1.5in, left=1.2in, right=1.2in]{geometry}
\usepackage{authblk}
\usepackage{amssymb, amsmath,framed}
\usepackage{graphicx}
\usepackage[round]{natbib}
\usepackage{float}
\usepackage{bm}
\usepackage[colorlinks=true,linkcolor=blue,urlcolor=blue,citecolor=blue,anchorcolor=blue]{hyperref}
\expandafter\ifx\csname package@font\endcsname\relax\else
 \expandafter\expandafter
 \expandafter\usepackage
 \expandafter\expandafter
 \expandafter{\csname package@font\endcsname}
\fi
\def\bq{\begin{equation}}
\def\eq{\end{equation}}
\def\bqy{\begin{eqnarray}}
\def\eqy{\end{eqnarray}}
\makeatother

\begin{document}

\title{The New Astronomical Frontier of Interstellar Objects}

\author{Amir Siraj\thanks{email: \texttt{amir.siraj@cfa.harvard.edu}}\,\,}

\author{Abraham Loeb\thanks{email: \texttt{aloeb@cfa.harvard.edu}}}
\affil{Department of Astronomy, Harvard University \\ 60 Garden Street, Cambridge, MA 02138, USA}

\date{}

\maketitle

\begin{abstract}
The upcoming commencement of the Vera C. Rubin Observatory’s Legacy Survey of Space of Time (LSST) will greatly enhance the discovery rate of interstellar objects (ISOs). `Oumuamua and Borisov were the first two ISOs confirmed in the Solar system, although the first interstellar meteor may have been discovered earlier. We discuss the properties of `Oumuamua and Borisov and explore the expected abundance of ISOs as a function of size in the solar neighborhood. We compare the expected abundance of ISOs to that of objects in the Oort cloud, and draw conclusions about the mass budget per star that is required to produce ISOs. We also investigate the possibility of ISOs being captured into bound orbits within the solar system, both from its birth star cluster and in the field. We examine the potential for ISOs to transport prebiotic or biotic material between planetary systems. We consider signatures of ISOs colliding with the Earth, the Moon, and neutron stars, as well as the possibility of differentiating ISOs from solar system objects in stellar occultation surveys. Finally, we discuss advantages that the imminent advent of LSST will afford the field of ISO studies, including large-number statistics that will reveal the origins of ISOs and discoveries of rare ISOs providing insights into exotic phenomena. One of the two branches of the newly established Galileo Project\footnote{\url{https://projects.iq.harvard.edu/galileo}} seeks to learn more about the nature of ISOs like `Oumuamua by performing new searches and designing follow-up observations.
\end{abstract}

\section{Introduction}

Interstellar objects (ISOs) are small bodies that are not gravitationally bound to any star. 1I/`Oumuamua was the first to be reported in the solar system in 2017 \citep{2017Natur.552..378M}, followed by 2I/Borisov in 2019 \citep{2020NatAs...4...53G}. CNEOS 2014-01-08 \citep{2019arXiv190407224S}, if confirmed, would be the first interstellar meteor larger than dust and the earliest detected ISO. The study of ISOs could serve as a novel lens through which planet formation and the prospects for life elsewhere in the Universe, among other astrophysical phenomena, could be studied \citep{1987AJ.....94.1330D, 2003Icar..166..141C, 2011MNRAS.417.2104V, 2014MNRAS.437.1127V, 2015PhyS...90f8001P, 2018ApJ...855L..10D, 2018ApJ...866..131M, 2018MNRAS.476.3031R,  2019MNRAS.490...21H, 2019arXiv190705427L, 2019AJ....157...86M, 2019ApJ...874L..34P,  2019ApJ...884L..22R, 2020ApJ...904L...4R, siraj_amir_2020_4075557,  2020ApJ...903L..20S, 2021ApJ...917...20H, 2021arXiv210811194L, 2021arXiv211015366M, 2021A&A...651A..38P, 2021MNRAS.507L..16S}.

Future observations of ISOs are of particular interest because all of the natural interpretations for the first ISO, `Oumuamua, invoked astrophysical objects we have never seen before \citep{2018ApJ...861...35R, 2019ApJ...872L..32M, 2019arXiv190500935S, 2019ApJ...885L..41F, 2020ApJ...896L...8S, 2020NatAs...4..852Z, 2021JGRE..12606706J, 2021ApJ...920...28S}. One of the most recent hypotheses, a nitrogen iceberg \citep{2021JGRE..12606706J, 2021JGRE..12606807D}, suffers from an untenable mass budget \citep{2021arXiv210314032S, 2021arXiv210811194L, 2021arXiv210904494P}. 

Specifically, in \citep{2021arXiv210314032S} we show that the fraction of heavy elements in stars that must be converted into exo-Plutos in order to explain the detection of `Oumuamua is,
\begin{align}
\begin{split}
    \label{eq:limit}
         & {f_{\star}}   \sim   \;  10^{3.3^{+0.75}_{-1.5}} \times \left[  \left( \frac{n}{0.1 \mathrm{\; AU^{-3}}}\right) \right. \times \\ & \left( \frac{m}{2.4 \times 10^{11} \mathrm{\; g}} \right) \left( \frac{\rho_{\star}}{0.04 \; \mathrm{M_{\odot} \; pc^{-3}}}\right)^{-1} \times \\ & \left( \frac{M_{\star} / (SFR \times \tau)}{10^2}\right) \left( \frac{D_{N2}}{3 \; \mathrm{km}}\right)^{-1} \times \\ &   \left( \frac{f_Z}{1.3 \%}\right)^{-1} \left( \frac{\epsilon}{0.2}\right)^{-1} \left. \left( \frac{\eta}{1}\right)^{-1} \left( \frac{f_{CR}}{10^2}\right) \right] \; , 
\end{split}
\end{align}
where $m = 2.4 \times 10^{11} \mathrm{\; g}$ is the initial mass of `Oumuamua \citep{2021JGRE..12606706J, 2021JGRE..12606807D}, $n = 0.1^{+0.457}_{-0.097} \mathrm{\; AU^{-3}}$ is the abundance of `Oumuamua-like objects \citep{2018ApJ...855L..10D}, $\rho_{\star} = 0.04 \; \mathrm{M_{\odot} \; pc^{-3}}$ is the local density of stellar mass \citep{2017MNRAS.470.1360B},  $(SFR \times \tau) / M_{\star} \sim 1\%$ is the fraction of the Galactic stellar population that `Oumuamua could have originated from where $M_{\star}$ is the Galactic stellar mass \citep{2015ApJ...806...96L, 2019ApJ...878L..11I, 2019A&A...624L...1M, 2021MNRAS.501..302A}, $D_{N2} \sim 3 \mathrm{\; km}$ is the $\mathrm{N_2}$ surface depth on Pluto \citep{1998ASSL..227..655C, 2016Natur.534...82M, 2021JGRE..12606807D}, $f_z$ is solar metallicity, $\epsilon \sim 0.2$ is the conservative dynamical fudge factor, $\eta \sim 1$ is the most conservative possible value for the fraction of the total mass in nitrogen icebergs that lies within a logarithmic bin around the inferred mass of `Oumuamua, $f_{CR} \sim 10^2$ is the cosmic-ray factor that specifies the required increase in the initial mass relative to the $\sim 0.1 \mathrm{ \; km}$ radius \citep{2021arXiv210904494P}, and where the quoted errors are the $95\%$ Poisson error bars for a single detection. This demonstrates that the the mass budget in exo-Pluto planets necessary to explain the detection of `Oumuamua as a nitrogen iceberg chipped off from a planetary surface requires a mass of heavy elements exceeding $10^6 \times$ the total quantity locked in stars with 95\% confidence, making the scenario untenable because only a small fraction of the mass in stars ends in exo-Plutos.

If `Oumuamua were a normal rock, the implied mass budget to produce such a population of objects would still be in tension with our current understanding of planetary systems \citep{2009ApJ...704..733M, 2018ApJ...866..131M, 2019AJ....157...86M}. Note that the mass budget discrepancy discussed here goes far beyond these previous tensions as `Oumuamua is interpreted to be composed of pure nitrogen. The mass budget is reduced considerably if Oumuamua is a thin object, $\lesssim 1 \mathrm{\; mm}$ \citep{2018ApJ...868L...1B}, in which case the total mass of heavy elements needed would be that of a few-kilometer scale asteroid per star \citep{loeb21}. The possibility of an artificial origin should be left on the table since all natural explanations for `Oumuamua, as of yet, invoke objects never seen before \citep{2021arXiv211015213L}. One of the two branches of the Galileo Project seeks to learn more about the nature of ISOs like `Oumuamua by performing new searches and designing follow-up observations.

In this review, we discuss the implications of ISOs for the solar system ($\S$\ref{imp}), in particular the required mass budget to produce ISOs ($\S$\ref{req}) and the exchange of ISOs with the local environment ($\S$\ref{exc}). We then explore novel methods of ISO discovery ($\S$\ref{nov}), including stellar occultations ($\S$\ref{ste}), interstellar meteors ($\S$\ref{int}), lunar impacts ($\S$\ref{lun}), and neutron star impacts ($\S$\ref{neu}). Finally, we consider some of the exciting possibilities that future observations of ISOs will hold ($\S$\ref{fut}).

\section{Implications of ISOs for the solar system}
\label{imp}
\subsection{Required mass budget}
\label{req}

The existence of ISOs imply a certain mass budget necessary for producing them, which may be informative as to their origins. We start with the detection of Borisov, the first confirmed interstellar comet with a known composition, which allowed for a calibration of the abundance for such a population \citep{2021MNRAS.507L..16S}, in parallel with the reasoning used in deriving the abundance of `Oumuamua-like objects \citep{2018ApJ...855L..10D}. Just like the abundance implied by `Oumuamua, Poisson uncertainties of several orders of magnitude dominate the error budget for Borisov-like objects. \cite{2021MNRAS.507L..16S} studied implications of the observational constraint on ISO abundance set by Borisov. Specifically, we investigated the relative abundances of interstellar and non-ISOs in the Oort cloud, explored the fraction of heavy metals locked in ISOs. 

\begin{figure}[hptb]
 \centering
\includegraphics[width=0.7\linewidth]{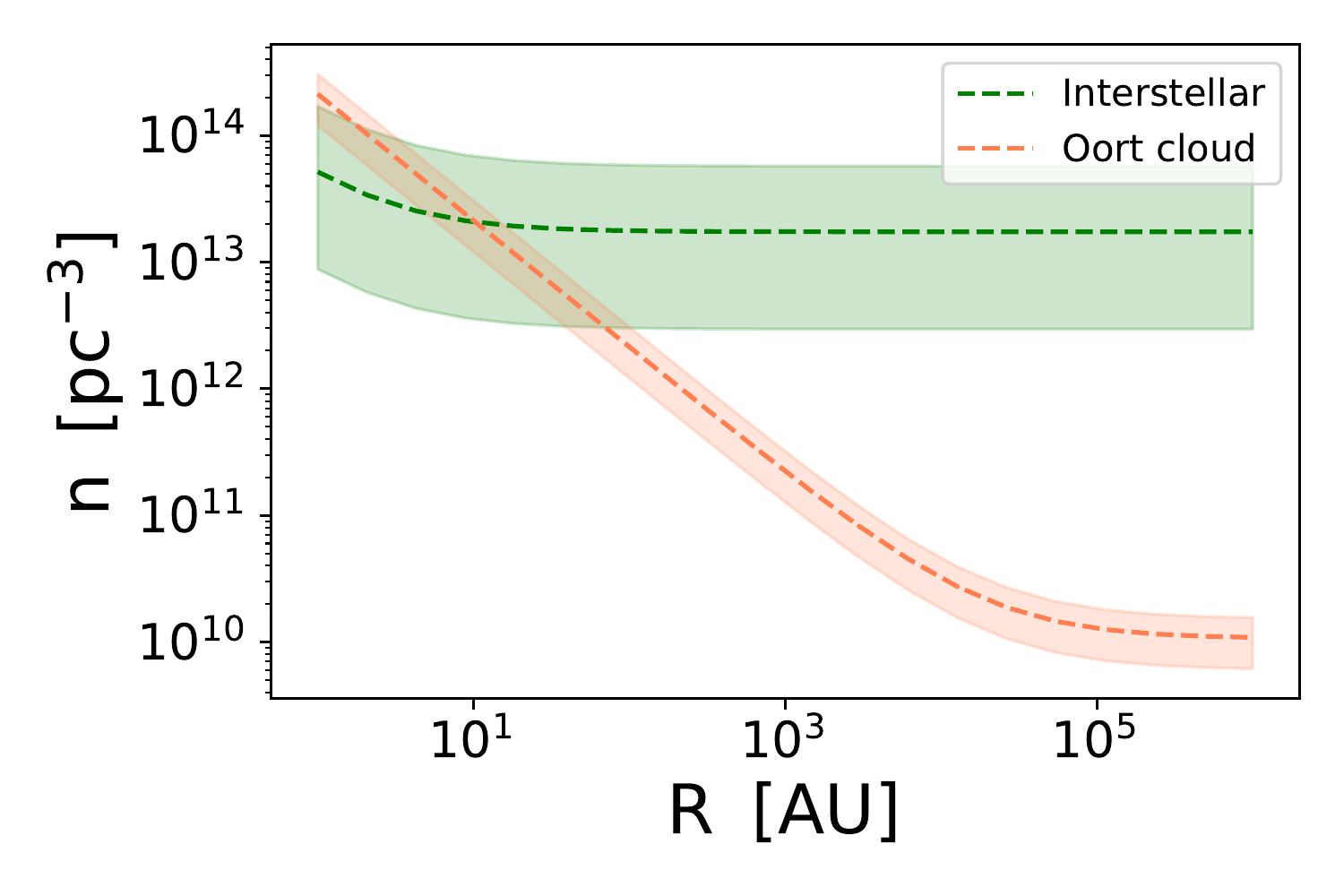}

\caption{Comparison of the local number density, $n$, of bound Oort cloud objects and ISOs as a function of distance, $R$, from the Sun due to gravitational focusing, for objects with diameters of $\sim 2.3 \mathrm{\; km}$ and adopting an ISO velocity dispersion at infinity of $\sim 30 \mathrm{\; km \; s^{-1}}$ \citep{2020ApJ...903L..20S} and a typical bound Oort cloud object semimajor axis of $\sim 2 \times 10^4 \mathrm{\; AU}$ \citep{2015SSRv..197..191D}. The bands indicate $1\sigma$ Poisson errors for the abundances of ISOs and bound Oort cloud objects.}
\label{fig:grav_focusing_vs_oc}
\end{figure}

Figure \ref{fig:grav_focusing_vs_oc} shows the primary result of \cite{2021MNRAS.507L..16S}: while ISOs are more numerous than bound Oort cloud objects in the Oort cloud ($R \sim 10^5 \mathrm{\; AU}$), bound Oort cloud objects are the dominant population in the inner Solar system, consistent with observations. Present observations of comets are confined to the vicinity of the Earth, a region of the Solar system where the characteristic orbital speed is $\sim 30 \mathrm{\; km \; s^{-1}}$. As a result, bound long-period comets, which have a low velocity dispersion of $\sim 0.3 \mathrm{\; km \; s^{-1}}$ at the outer envelope of the Oort cloud, experience an enhancement by gravitational focusing in their density of order $\sim 10^4$, whereas ISOs, which have a higher background velocity dispersion of $\sim 30 \mathrm{\; km \; s^{-1}}$, only experience an enhancement in density of order unity relative to their abundance far from the Sun. Given that the number density of ISOs may be $\sim 10^3$ larger than that of bound Oort cloud objects far from the Sun, the Oort cloud objects may be still a factor of $\sim 10$ more abundant than ISOs in the inner Solar system, due to the unequal influence of gravitational focusing on the two populations. The fact that ISOs outnumber Oort cloud objects per star is consistent with the Oort cloud having lost most of its initial mass. However, the degree to which ISOs outnumber Oort cloud objects is still very uncertain. Stellar occultation surveys of the Oort cloud will be capable of confirming the results presented here, by differentiating between the two populations through speed relative to the Sun \citep{2007AJ....134.1596N, 2020ApJ...891L...3S} . 

Regarding the fraction of heavy metals locked in ISOs, the \cite{2021MNRAS.507L..16S} results indicate that the fraction of carbon and oxygen contained in ISOs is sensitive to the size distribution of ISOs, varying by nearly four orders of magnitude depending on the parameters chosen for the broken power-law size distribution consistent with the detection of Borisov. Bounds of $\pm 0.7$ about the scale invariant value of $q = 3$ are necessary for the mass of heavy elements (metals) contained in ISOs not to exceed that of stars or the ISM. Interestingly, the most conservative power-law fit of $q = 3$, representing a scale-free distribution, implies that nearly $\sim 1 \%$ of the carbon and oxygen contained in stars and the ISM is locked in ISOs. This can therefore be regarded as a lower bound on the actual fraction of the Galactic metals budget in ISOs, and is physically interesting because it is comparable to metals budget of the minimum mass solar nebula model (MMSN) \citep{1977MNRAS.180...57W, 1981PThPS..70...35H, 2007ApJ...671..878D, 2009ApJ...698..606C} -- suggesting that if ISOs are formed in protoplanetary disks a significant proportion of protoplanetary material is ejected during the planetary formation process. Additionally, if some ISOs have a finite lifetime, a larger mass budget is required. This presents a major challenge for recent planetary system formation simulations, which result in an ejected object mass density several orders of magnitude too low to be consistent with this constraint \citep{2021A&A...647A.136P}. This conclusion is consistent with the difficulty in reconciling the implied abundance of ISOs with planetary system origins \citep{2018ApJ...866..131M, 2019AJ....157...86M}.

\begin{figure}[hptb]
 \centering
\includegraphics[width=0.7\linewidth]{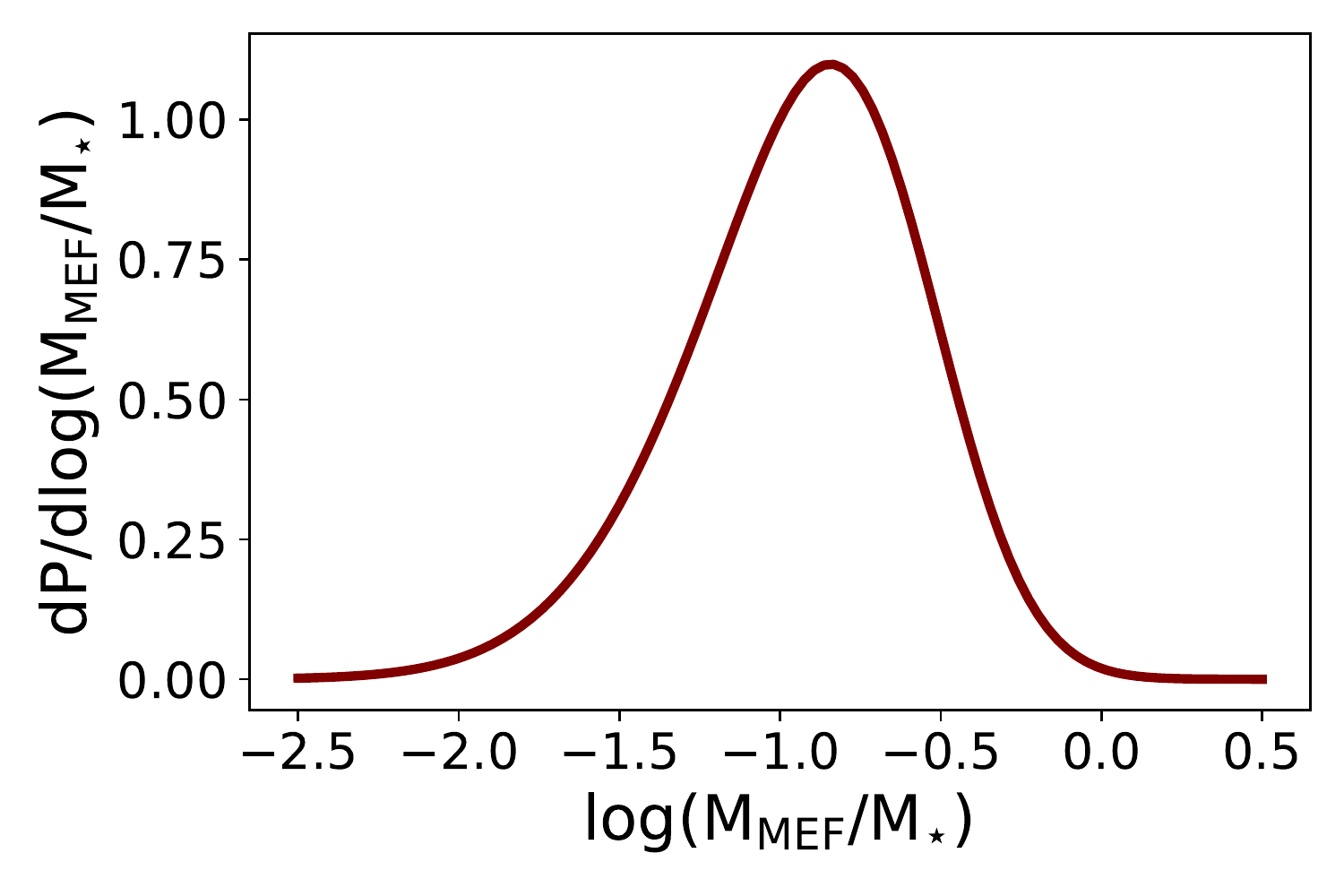}
\caption{Normalized probability distribution for $(M_{MEF} / M_{\star})$, the mass fraction per star necessary to generate the observed of population of ISOs larger than `Oumuamua.}
\label{fig:MEF}
\end{figure}

Estimates of the mass contained in ISOs can be improved by including `Oumuamua. \cite{2021arXiv210813429S}  employed a Monte Carlo simulation incorporating the probability distributions for size and abundance for `Oumuamua-like objects, Borisov-like objects, and rocky rogue planets. The model outputs a probability distribution for the mass locked in ISOs in size range $\sim 10^{4} - 10^{9} \; \mathrm{cm}$. How much mass per star is necessary to form the population of ISOs larger than `Oumuamua, and how does this compare to the MMSN? Given the Sun's metallicity of 1.4\% \citep{2009ARA&A..47..481A}, and since `Oumuamua-like objects, Borisov-like objects, and rocky rogue planets are all primarily composed of elements heavier than hydrogen and helium by mass: the total mass budget necessary to produce this population of interstellar obejcts, which we dub the Minimum Ejection Fraction (MEF), is related to the mass budget of ISOs as follows, $M_{MEF} = (M_{ISO} / 1.4\%)$. Given the Monte Carlo simulation defined above, we find that the central value to be $M_{MEF} \sim 0.1 M_{\star}$, with a 95\% confidence interval of $0.02 M_{\star} - 0.5 M_{\star}$. Figure \ref{fig:MEF} shows the overall probability distribution for $M_{MEF} / M_{\star}$.

The primary implication of these results is that the amount of mass needed to form ISOs larger than `Oumuamua is a substantial fraction of host star mass, between 2\% and 50\%. This provides a new constraint on planetary system formation, since it requires an order of magnitude more mass than the MMSN \citep{1977MNRAS.180...57W, 1981PThPS..70...35H, 2007ApJ...671..878D, 2009ApJ...698..606C}. These results suggest a highly efficient route for converting protostellar matter into $\gtrsim 0.1 \mathrm{\; km}$ planetesimals and for ejecting them from their parent stars, and changes the paradigm regarding observational constraints on the planetary system formation process.

Additionally, these results imply that the ejected mass from stars exceeds the retained mass. Neither the mass budget of the solar system's protoplanetary disk \citep{2013ApJ...768...45N, 2014ApJ...792..127R, 2017AJ....153..153D} nor the observed protoplanetary or debris disks around other stars could provide sufficient material for the formation of ISOs \citep{2020ARA&A..58..483A}. Large planetesimals must be ejected by dynamical processes, and such processes must provide kicks strong enough to unbind them from their parent stars. Stellar binaries are likely to eject the inner region of protoplanetary disks \citep{2018ApJ...866..131M, 2018MNRAS.478L..49J} out to approximately twice their semimajor axis for disks larger than the binary separation \citep{2011ASL.....4..181C, 2020ApJ...900...43T}. Giant planets could create a ring/gap substructure in their protoplanetary disks by ejecting planetesimals from their vicinity \citep{2019ApJ...884L..22R}. The origins of ISOs can be inferred through their velocity distribution, once a sufficient number of them have been detected \citep{2020ApJ...903L..20S}.

Although it was suggested that ISOs may seed planet formation \citep{2019ApJ...874L..34P, 2019MNRAS.487.3324G, 2021arXiv211015366M}, we note that if ISOs are made in protoplanetary disks, then their abundance upon formation overwhelms that of ISOs in the field because the volume and lifetime of protoplanetary disks are much smaller than the volume and crossing time of interstellar space.

\subsection{Exchange with local environment}
\label{exc}

ISOs facilitate exchange between the solar system and its local environment. \cite{2019ApJ...872L..10S} suggested that ISOs could theoretically be captured in the solar system a result of restricted three-body interactions with the Sun and Jupiter, and presented preliminary simulations of strong capture scenarios. \cite{2020MNRAS.493L..59H} performed a similar calculation for the weaker capture, which is more likely. \cite{2020MNRAS.497L..46M} showed that there is no evidence of trapped ISOs in the solar system at present. \cite{2021PSJ.....2...53N} and \cite{2021arXiv210911017N} generalized results for ISO capture, and showed that they should be very rare, and therefore difficult to find, in the solar system. However: while the solar system does not capture and retain ISOs at a significant rate in its current configuration, it could have captured numerous objects in the past if it was born with a stellar twin in a binary system.

Simulations of outer Oort cloud (OOC)\footnote{Defined here as the collection of solar system bodies with semimajor axes of $\sim 10^4 - 10^5 \mathrm{\; AU}$, and with orbits decoupled from Neptune.} formation in the Solar system  \citep{2004come.book..153D, 2008Icar..197..221K, 2010A&A...516A..72B} have difficulties reproducing the observed ratio between scattered disk) (SD)\footnote{Defined here as the collection of solar system bodies with semimajor axes of $\lesssim 10^3 \mathrm{\; AU}$, and with orbits controlled by Neptune.} and OOC objects \citep{1997Sci...276.1670D, 2008AJ....136.1079L} As a result, the origin of the OOC is an unsolved puzzle. Scenarios positing that the formation of the OOC occurred in the stellar birth cluster of the Sun tend to rely on drag from dense cluster gas \citep{2000Icar..145..580F, 2006Icar..184...59B,  2008Icar..197..221K, 2010Sci...329..187L, 2012Icar..217....1B}, a factor that hinders the scattering of comets to large distances, reducing the plausibility of the explanations \citep{2007Icar..191..413B, 2013Icar..225...40B}. There are $N_{OC} \sim 7.6 \pm 3.3 \times 10^{10}$ OOC bodies and $N_{SD} \sim 1.7^{+ 3.0}_{- 0.9} \times 10^9$ SD bodies with diameters of $D > 2.3 \mathrm{\; km}$ \citep{2013Icar..225...40B, 2017A&A...598A.110R}. Simulations of OOC formation due to a dynamical instability in the solar system result in an OOC/SD ratio of $N_{OC}/N_{SD} \sim 12 \pm 1$, which is in tension, but not incompatible with, the observed ratio \citep{2013Icar..225...40B}. Separately, clustering of extreme trans-Neptunian objects (ETNOs) in the outer solar system suggest the possible existence of a planet, labeled Planet Nine, at a distance of $\sim 500 \mathrm{\; AU}$ from the Sun \citep{2016ApJ...824L..23B, 2019PhR...805....1B}. \cite{2020AJ....160...50Z} argued that Planet Nine may not exist, and its observed gravitational effects could potentially be caused by an unobserved ring of small bodies in the outer solar system. There also exists the possibility that the clustering is a statistical fluke \citep{2020AJ....159..285C}. The origin of Planet Nine, if it exists, is a second unsolved puzzle in the outskirts of the Solar system. Possible solutions include \citep{2019PhR...805....1B} formation amongst the giant planets followed by scattering and orbital circularization \citep{2006Icar..184...59B, 2012Icar..217....1B, 2016ApJ...823L...3L}, and capture in the solar birth cluster \citep{2016ApJ...823L...3L, 2016MNRAS.460L.109M, 2017MNRAS.472L..75P}.

\begin{figure}[hptb]
\smallskip
  \centering
  \includegraphics[width=0.7\linewidth]{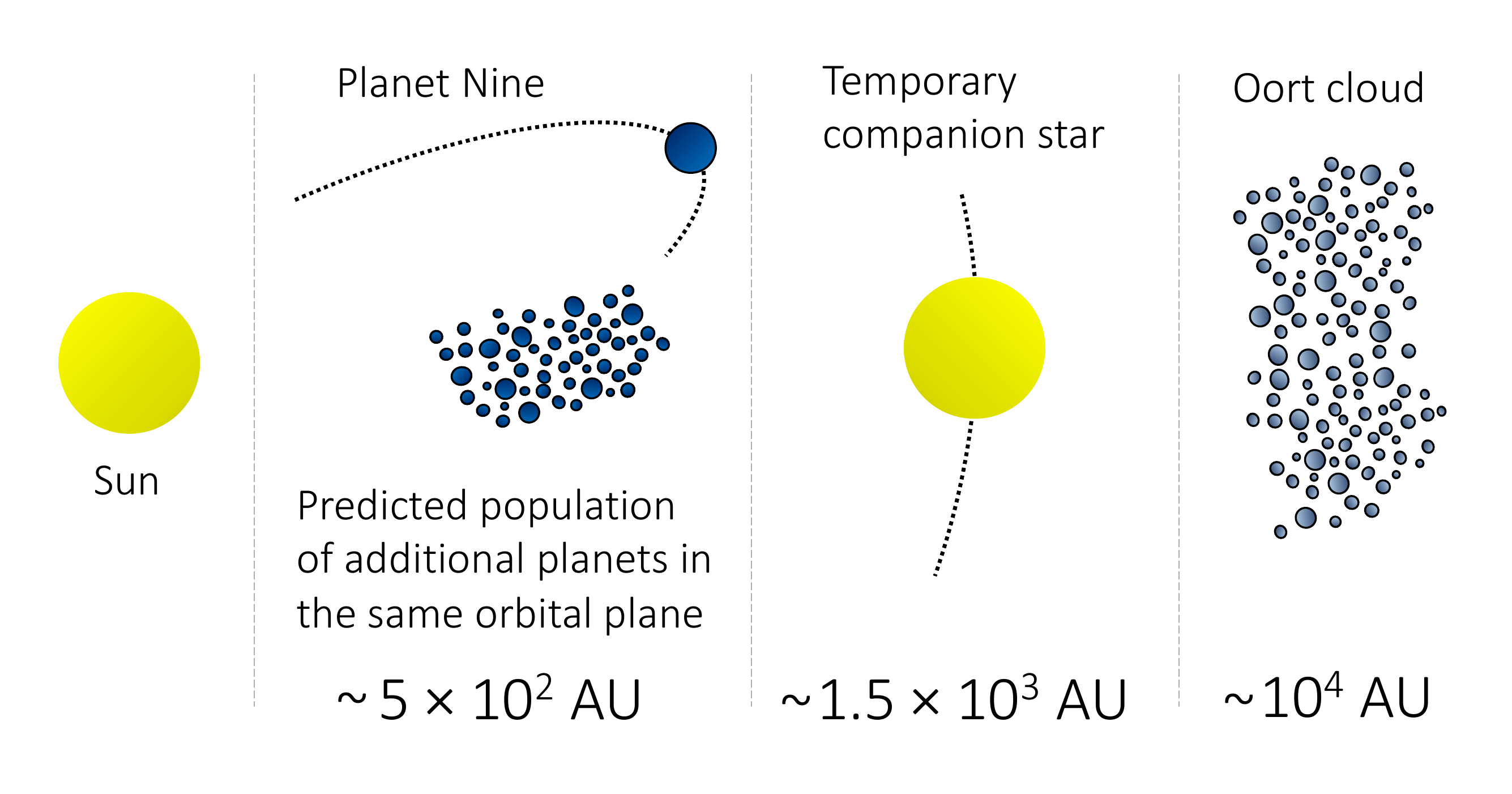}
    \caption{Sketch of scenario considered in \cite{2020ApJ...899L..24S} (not to scale).
}
    \label{fig:sketch}
\end{figure} 

Interestingly, stellar binary systems are capable of capturing background objects via three-body processes \citep{1975MNRAS.173..729H, 1983Obs...103....1V}, leading to capture rates that are enhanced relative to lone stars \citep{2018ApJ...868L..12G, 2020arXiv200102235S}. Current binary companions to the Sun were previously considered \citep{2005EM&P...97..459M, 2010MNRAS.407L..99M}, and subsequently ruled out \citep{2014ApJ...781....4L}. In \cite{2020ApJ...899L..24S}, we consider a temporary binary companion to the Sun that could have existed only in the solar birth cluster, and explore the plausibility and implications of such a possibility for both the formation of the OOC and the capture of Planet Nine. Specifically, we propose that an equal-mass binary companion to the Sun in the solar birth cluster at a separation of $\sim 10^3 \mathrm{\; AU}$ would explain the formation of the observed population of OOC objects and the putative existence of Planet Nine. Separations greater than the fiducial example given here, $a \sim 1500 \; \mathrm{AU}$, are entirely plausible; the capture cross-section would simply scale as $a^{-1}$ and the likelihood of ejection in the birth cluster as $a^{1/2}$, up to a maximum of $a \sim 6000 \; \mathrm{AU}$ since the chance of ejection in the birth cluster would then be of order unity. If Planet Nine is discovered, evidence of a captured origin, as opposed to formation within the Solar system, could potentially come from a cloud of objects with associated orbits \citep{2016MNRAS.460L.109M}. Accounting for the likelihood of the binary configuration considered here, the discovery of a captured Planet Nine would favor our binary model by a factor of $\sim 10$, when the increased likelihoods of both forming the OOC and capturing Planet Nine are considered.

The specific smoking gun for our binary model will be a significant overabundance of dwarf planets with similar orbits to Planet Nine, since the the capture cross-section for such objects would have been a factor of $\sim 20$ larger than implied by the conventional lone stellar model, and given that orbits situated closer to the proposed binary than Planet Nine would be unstable \citep{2020AJ....159...80Q}. These objects could potentially be detected by the Legacy Survey of Space and Time (LSST)\footnote{\url{https://www.lsst.org/}} on the Vera C. Rubin Observatory. In addition, since the binary model would bring the likelihood of Planet Nine capture in the solar birth cluster near unity, the existence of a captured planet in addition to Planet Nine would be probable. Detailed modeling of the effects of a binary on long-period comets, the solar obliquity, and ETNOs will allow for the development of additional tests.

In addition to playing a potentially critical role in the formation of the Oort cloud, ISOs could have also had effects closer to home over the past 4.5 Gyr. Specifically, ISOs could have carried atmospheric samples of prebiotic or possibly biotic material from Earth, to other star systems. Panspermia is the idea that life can propagate from one planet to another \citep{2010SSRv..156..239W, 2010IJAsB...9..119W}. Impacts on the surface of a planet can launch debris at above the escape speed of the planet, thereby allowing debris spread throughout the planetary system and constituting a plausible mechanism for exchanging life between planets orbiting the same star \citep{2005M&PS...40.1393F, 2000Icar..145..391M}. However, it is difficult to eject life-bearing material at speeds above the escape speed from a planetary system that is effectively shielded from destructive radiation, presenting a significant challenge for spreading life between stars \citep{2010SSRv..156..239W}. Life in the Earth's atmosphere has been detected up to an altitude of 77 km \citep{imshenetsky1978upper}, constituting a reservoir of microbes that objects grazing the atmosphere could draw from. Here, we discuss the potential for three classes of objects to carry out such a process: objects that were on originally hyperbolic, parabolic, and elliptical orbits, respectively.

First, we consider the cases of objects on originally hyperbolic or parabolic orbits, motivated from a theoretical standpoint. Long-period comets (LPCs) represent a population of bodies that can easily be ejected from the Solar System by gravitational interactions with planets due to their low gravitational binding energies and planet-crossing orbits. This makes them ideal, in principle, for both picking up life from Earth and exporting it out of from the Solar System.

The high speed and abundance of ISOs (ISOs) make them, as well as LPCs, potential exporters of life from Earth to exoplanetary systems. \cite{2018ApJ...868L..12G} and \cite{2018AJ....156..193L} demonstrate dynamically that ejected objects can be gravitationally captured by other star systems. In \cite{2020IJAsB..19..260S}, we study whether it is possible for ISOs and LPCs to have exported life from Earth's atmosphere out of the solar system. We estimate the total number of exportation events over the lifetime of the Earth to be $\sim 1 - 10$ for LPCs and $\sim 1 - 50$ for ISOs. If life existed above an altitude of 100 km, we find that up to $\sim 10^5$ exportation events could have occurred over Earth's lifetime.  An important comparison to make is to the conventional mode of panspermia involving impacts and subsequent ejecta. \cite{2012AsBio..12..754B} find that $10^{14} - 3 \times 10^{16}$ objects with mass $> 10 \mathrm{\; kg}$ were transferred from the Sun to its nearest neighbors in the birth cluster. Assuming a density of $600 \; \mathrm{kg \; m^{-3}}$ and a cumulative size distribution exponent of $-1.5$ in terms of size, we find that $\sim 10^8 - 10^{10}$ km-sized objects were transferred. If $\gtrsim 10^{-6}$ of such objects had viable microbes \citep{2012AsBio..12..754B}, this would yield $> 10^2 - 10^4$ objects, which is higher than the number of LPCs and ISOs capable of transferring life, given the cutoff height of 80 km. Improved measurements of the size distribution of LPCs and ISOs would allow for more precise estimates. 

We now consider the cases of objects on originally elliptical orbits, motivated by the discovery of a small earth-grazing object in 2017. Specifically, \citet{2020AJ....159..191S} reported a detection by the Desert Fireball Network (DFN) \footnote{\url{http://fireballsinthesky.com.au/}} of a $\sim$30-$\mathrm{cm}$ object that reached a minimum altitude $\sim$58.5 $\; \mathrm{km}$ in the Earth's atmosphere during a $90 \; \mathrm{s}$ grazing event, and~was transferred from an Apollo-type orbit into a Jupiter-family comet orbit, making it likely to be ejected from the Solar System during a future gravitational encounter with Jupiter. This detection represents a new class of objects that can pick up life in the Earth's atmosphere before being ejected from the Solar System: rocky, inner-Solar System bodies that are scattered into Jupiter-crossing orbits after their grazing interactions with Earth and are subsequently ejected. These~objects are important, since, based on their characteristically higher densities relative to icy, outer-Solar System bodies, they can survive passes through the Earth's atmosphere at significantly smaller~sizes.

\begin{figure}[H]
  \centering
  \includegraphics[width=0.7\linewidth]{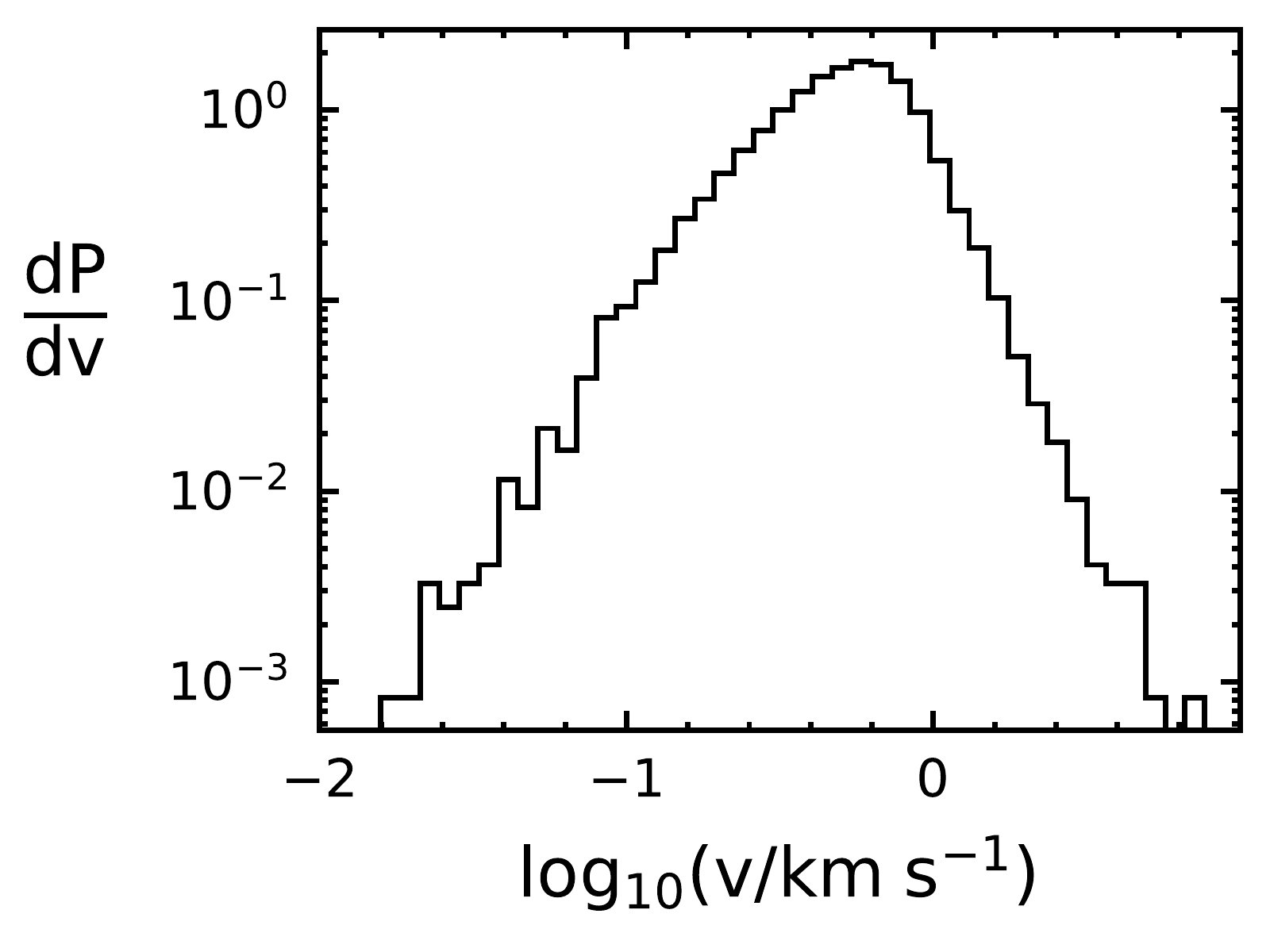}
    \caption{Distribution of relative speed at capture between $1.9 \times 10^4$ simulated captured objects and their respective binary systems. P is probability and v is the relative speed at capture.
}
    \label{fig:fig2}
\end{figure}

It has been demonstrated that many microbes can likely survive the accelerations associated with transfer onto the grazing object \citep{2001E&PSL.189....1M, 2011PNAS..108.7997D}, although~specific detections in the atmosphere linked to microbes that have undergone such laboratory testing have not yet been~made. Polyextremophiles such as \textit{Deinococcus radiodurans} are estimated to die on an exponential timescale of $\sim10^5 \; \mathrm{years}$ with minimal radiation shielding \citep{2018ApJ...868L..12G, phillips2012physical, 2004IJAsB...3...73B, 2000Icar..145..391M}. While the abundance of polyextremophiles in the atmosphere is yet unconstrained, in~our analysis, we assume a fiducial estimate of 1 polyextremophile colony ($\lesssim$10\textsuperscript{$-$4} of incident microbe colonies) per transporting object. The~interiors of objects with diameters greater than 20 cm are not expected to be heated to more than $100~^{\circ}$C during a passage through the atmosphere \citep{2004MNRAS.348...46N}, and~many asteroids are known to have significant porosity \citep{2000Icar..146..213B, 2006LPI....37.2214B},  thus  we adopt a model in which incident microbes become lodged inside the object and shielded from the exterior heating. We note that this assumption is not critical; increasing the minimum altitude by a few scale heights above the observed altitude of the \cite{ 2020AJ....159..191S} bolide would result in minimal heating ($<$100~\textsuperscript{$\circ$}C) while only reducing the atmospheric cross section by a small factor, potentially allowing for microbes on the surface in the case of non-porous~objects. In \cite{2020arXiv200102235S}, we study the likelihood of life-bearing Solar System bodies being captured by exoplanetary systems. Using a Monte Carlo simulation, we calculated the likelihood of potentially life-bearing Solar System bodies to get captured by exoplanetary systems. The~total number of objects captured by exoplanetary systems over the lifetime of the Solar System is $10^{7}$--$10^{9}$, with~the total number of objects with the possibility of living microbes on them at the time of capture estimated to be $10$--$1000$. The~distribution of relative speed between the object and the binary system for the instances resulting in capture is shown in Figure \ref{fig:fig2}, indicating that the highest chance of capture is achieved for relative speeds in the range $0.1 - 1 \mathrm{\; km \; s^{-1}}$.

Panspermia is a multi-stage process and not all steps are addressed here, most notably the delivery of microbes to a planet within binary systems. Many of the probability estimates derived here are related to explicitly stated fiducial values, and~are easily scaled to accommodate future research. Further detections of Earthgrazing objects transferred into Jupiter-crossing orbits will allow for more precise estimates of this process. Measurements of the abundance of polyextremophiles in the atmosphere will also refine the estimates reviewed here.  Additional studies of polyextremophiles in space will advance the biological understanding of this and other panspermia~channels. 

\section{Novel methods of ISO discovery}
\label{nov}

In addition to searching for ISOs via the sunlight they reflect, there are several other observable signatures that these rare objects should exhibit. Here, we review some of these possibilities, including stellar occultations as well as collisions for ISOs with the Earth, the Moon, and neutron stars.

\subsection{Stellar occultations}
\label{ste}

First, we consider stellar occultations by ISOs. Transient diffraction patterns caused by the occultation of a distant star due to an intervening small body have been used to search for Kuiper Belt objects (KBOs) and Oort Cloud objects (OCOs) in the Solar System \citep{1976Natur.259..290B, 1992QJRAS..33...45D, 2000Icar..147..530R, 2007AJ....134.1596N, 2009Natur.462..895S, 2012ApJ...761..150S, 2019NatAs...3..301A}. In \cite{2020ApJ...891L...3S}, we propose an analagous search for ISOs, flagged by their unusual inclinations and unique kinematics leading a distribution of characteristic distribution of durations, the peak of which lies between that of Kuiper belt objects and that of Oort cloud objects. The data from such a search would calibrate population parameters for ISOs crucial for constraining formation theories of exoplanetary systems \citep{1987AJ.....94.1330D, 2003Icar..166..141C, 2011MNRAS.417.2104V, 2014MNRAS.437.1127V, 2015PhyS...90f8001P, 2018ApJ...855L..10D, 2018MNRAS.476.3031R,  2019MNRAS.490...21H, 2019ApJ...874L..34P, 2021NewA...8401545S}.

\begin{figure*}[!th]
  \centering
  \includegraphics[width=.7\linewidth]{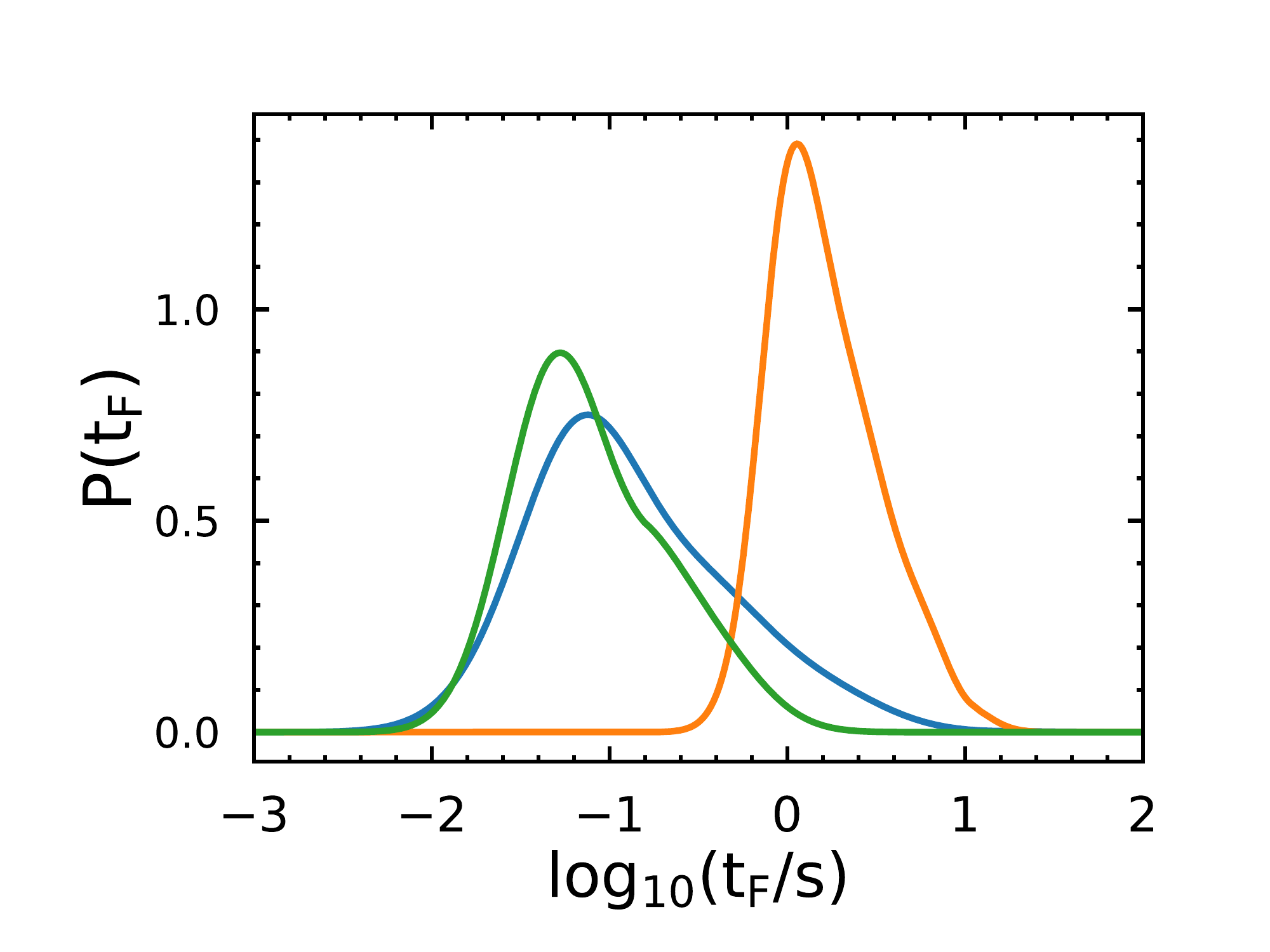}
    \caption{Expected probability distribution of Fresnel crossing times during stellar occultations observed from Earth for ISOs (blue), KBOs (green), and OCOs (orange), for a R = 12 solar-type star.}
    \label{fig:fig3}
\end{figure*}

To determine how easily discernable ISO, KBO, and OCO occultation signatures are from each other, and to understand the distribution of timescales on which ISO occultations occur, we numerically simulated the distribution of expected Fresnel distance crossing times for each population, assuming a R = 12 solar-type star. The resulting distributions of $t_F$ are shown in Figure \ref{fig:fig3}. We find that the timescales for ISO and OCO occultation events are distinct. The ISO timescale distribution peaks at $t_F = 0.1 \; \mathrm{s}$, and a survey can avoid KBOs by pointing away from the plane of the ecliptic. There are $\sim 7 \times 10^6$ stars in a with magnitude $R \lesssim 12.5$ \citep{2018RNAAS...2...51M}. An all-sky network of 1-m telescopes continuously monitoring all R = 12 stars with a time resolutions of $0.1 \; \mathrm{s}$ should yield a discovery rate of $\sim 1$ ISO per year. This would be a significant improvement on the current discovery rate of an ISO every few years. Our method supplements direct detection through reflected sunlight but nearby events could benefit from both methods of detection. By measuring each occultation in two or three colors, the radius and distance of the occulting object can be constrained \citep{1992QJRAS..33...45D}. The results of such a survey would provide invaluable new information on the size distribution, composition, and possible origin of ISOs. Such information would be particularly valuable given the puzzle of the first two confirmed ISOs, `Oumuamua and Borisov, having such different physical characteristics \citep{2017Natur.552..378M, 2020NatAs...4...53G}, and given the high implied abundance of ISOs relative to previous predictions \citep{2009ApJ...704..733M}.

\subsection{Interstellar meteors}
\label{int}

We next consider interstellar meteor collisions with the Earth, as discussed in \cite{2019arXiv190603270S}. Small ISOs can strike the Earth and appear as meteors in the sky. There is significant evidence for previous detections of interstellar micrometeors \citep{1993mtpb.conf...53B, 1994A&A...288..330H, 1996Natur.380..323T, 2000JGR...10510353B, 2001GeoRL..28.1929M, 2002ApJ...567..323M, 2004EM&P...95..221W, 2007AstBu..62..301A, 2012ApJ...745..161M, 2017AJ....153..133E, 2019msme.book..235H, 2019arXiv190407224S}. CNEOS 2014-01-08 is tentatively the first interstellar meteor larger than dust discovered in the Solar System \citep{2019arXiv190407224S}. Spectroscopy of gaseous debris of interstellar meteors as they burn up in the Earth's atmosphere could reveal their composition \citep{2019arXiv190407224S}. The cores of meteoroids with radii larger than $\sim 5$cm can reach the ground in the form of meteorites \citep{LIPSCHUTZ2014625}. Additionally, meteoroids on smaller size scales could be accelerated from the Poynting-Robertson effect and could have potential origins in the ISM. Hence, interstellar meteors above this size are optimal for a systematic study of physical extrasolar material (in addition to the spectroscopy of the hot gases as the meteor burns up). Since we expect interstellar meteors of this size to strike the Earth a few times per year, a network of steerable telescopes monitoring the sky above all land on Earth could detect an interstellar meteor of this size every few years.

\begin{figure*}
  \centering
  \includegraphics[width=0.7\linewidth]{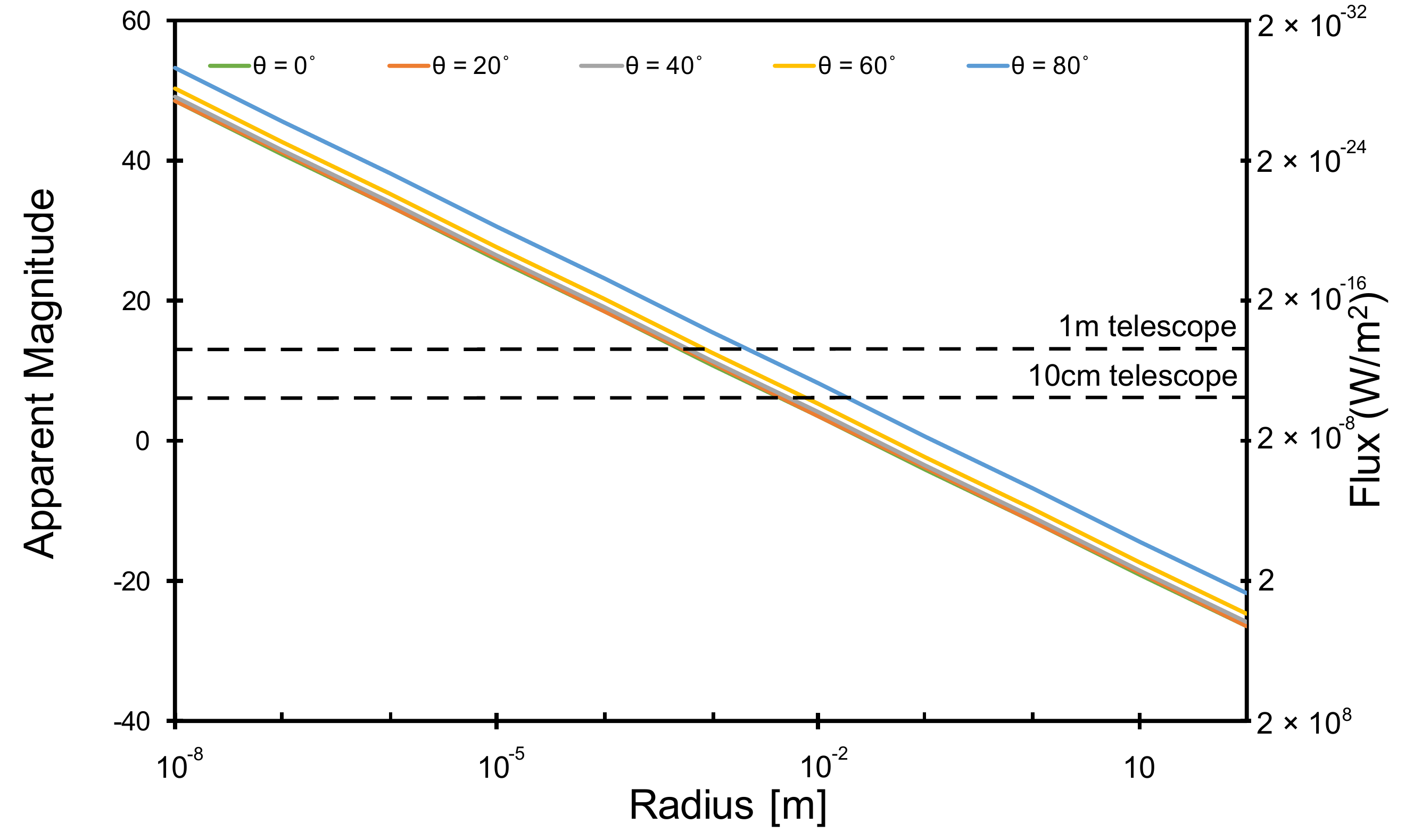}
    \caption{Apparent visual magnitude for $5 \times 10^{-4}$ s exposures as a function of meteor size at an altitude of 100 km, with $\theta$ measured from zenith to horizon. We assume a fiducial density of $\rho = 2\times 10^3 \;\mathrm{kg \; m^{-3}}$, an impact speed of $v_{impact} = 30 \; \mathrm{km \; s^{-1}}$, and that the meteor lifetime is on the order of a few seconds. Dashed lines the flux limit for a 10-$\sigma$ detection in each spectral bin with a spectral resolution of $R=10^3$, for telescope apertures of diameter 10cm and 1m.}
    \label{fig:2}
\end{figure*}

The maximum height for $r \sim 5 \times 10^{-2}\; \mathrm{m}$ meteors is $\sim 70$km \citep{2005M&PS...40..817C}, so their lifetimes range from $\sim 0.5 \; \mathrm{s}$ to $\sim 3 \mathrm{\; s}$, depending on impact velocity and angle. The ablation vapor cools within a few seconds before recombination occurs \citep{Jenniskens2015}. It is therefore imperative that the alert, the slewing of the relevant telescope, and the spectroscopy take place within $\sim 1$s, which requires the capability of slewing to any point on the sky within $\sim 0.5$s. Figure~\ref{fig:2} shows brightness as a function of size. 10cm telescopes with field of view 10 square degrees, operating at a frame rate of 2000 fps (fast enough to determine the meteor speed to $\sim \pm 10^{-2} \; \mathrm{km\;s^{-1}}$) can achieve 10-$\sigma$ detection in each spectral bin with a spectral resolution of $R=10^3$ for meteors down to size $r \sim 2 \times 10^{-2} \; \mathrm{m}$. Any system of camera-equipped satellites that constantly monitor Earth, such as those operated by the U.S. Government that provided the CNEOS bolide catalog\footnote{\url{https://cneos.jpl.nasa.gov/fireballs/}}, could notify within $\sim 0.1$s (the travel time for light between geosynchronous orbit and Earth) the nearest steerable telescope about an incoming meteor with excess heliocentric velocity as it enters the atmosphere at $\sim 100 \mathrm{\; km}$. A conservative estimate for the total area of $\sim 70 \mathrm{\; km}$ altitude atmosphere visible from a system of two telescopes separated by $100 \mathrm{\; km}$ is $5 \times 10^5 \; \mathrm{km^2}$, so to cover all land on Earth would require $\sim 300$ systems, or $\sim 600$ total steerable 10cm telescopes capable of slewing to any point on the sky within $\sim 0.5$s. The systems would be similar to CAMO but with an all-sky field of view, like AMOS and CAMS\footnote{\url{http://cams.seti.org/}} \citep{2013Icar..225..614W, 2015P&SS..118..102T}. We therefore advocate for an global alert system based on existing satellite infrastructure, combined with an associated network of steerable 10cm telescopes, to conduct real-time remote spectroscopy of the hot gases as $r \geq 5 \times 10^{-2}\; \mathrm{m}$ interstellar meteors burn up, and to precisely determine their trajectories for the immediate retreival of interstellar meteorite samples. The needed infrastructure may already exist, including networks such as the Las Cumbres Observatory.\footnote{\url{https://lco.global/}}

\subsection{Lunar impacts}
\label{lun}

Small ISOs should also collide with the Moon. Whereas on Earth small meteoroids burn up high in the atmosphere, on the Moon they directly impact the surface, generating bright flashes of light and craters \citep{2000ApJ...542L..65B, 2010EM&P..107...55B, GOEL201554, 2015MNRAS.454..344O, 2006Icar..184..319O, 2019MNRAS.484.5212A}. In \cite{2020AcAau.173...53S}, we propose a mission dedicated to the discovery and characterization of interstellar impacts on the Moon. We show that a lunar satellite equipped with a telescope to monitor such events in real time could allow for precise velocity, mass, and density determination of ISOs. Immediate follow-up spectroscopy as well as exploration of the resulting craters could reveal detailed composition of the impactors. Such data would calibrate population parameters for ISOs crucial for constraining theories of planetary formation and for assessing the habitability of exoplanets \citep{1987AJ.....94.1330D, 2003Icar..166..141C, 2011MNRAS.417.2104V, 2014MNRAS.437.1127V, 2015PhyS...90f8001P, 2018ApJ...855L..10D, 2018MNRAS.476.3031R,  2019MNRAS.490...21H, 2019ApJ...874L..34P, 2021NewA...8401545S}, as well as the search for the building blocks of extraterrestrial life \citep{2019arXiv190705427L}. Such a satellite could also serve as a laboratory for hypervelocity impacts \citep{2000ApJ...542L..65B, 2001ESASP.495..533D, 2010EM&P..107...55B, 2015MNRAS.454..344O, 2006Icar..184..319O, GOEL201554, 2015JGRE..120.1847S, 2019MNRAS.484.5212A}, constraining physical processes difficult to replicate on Earth.

\begin{figure}
  \centering
  \includegraphics[width=.7\linewidth]{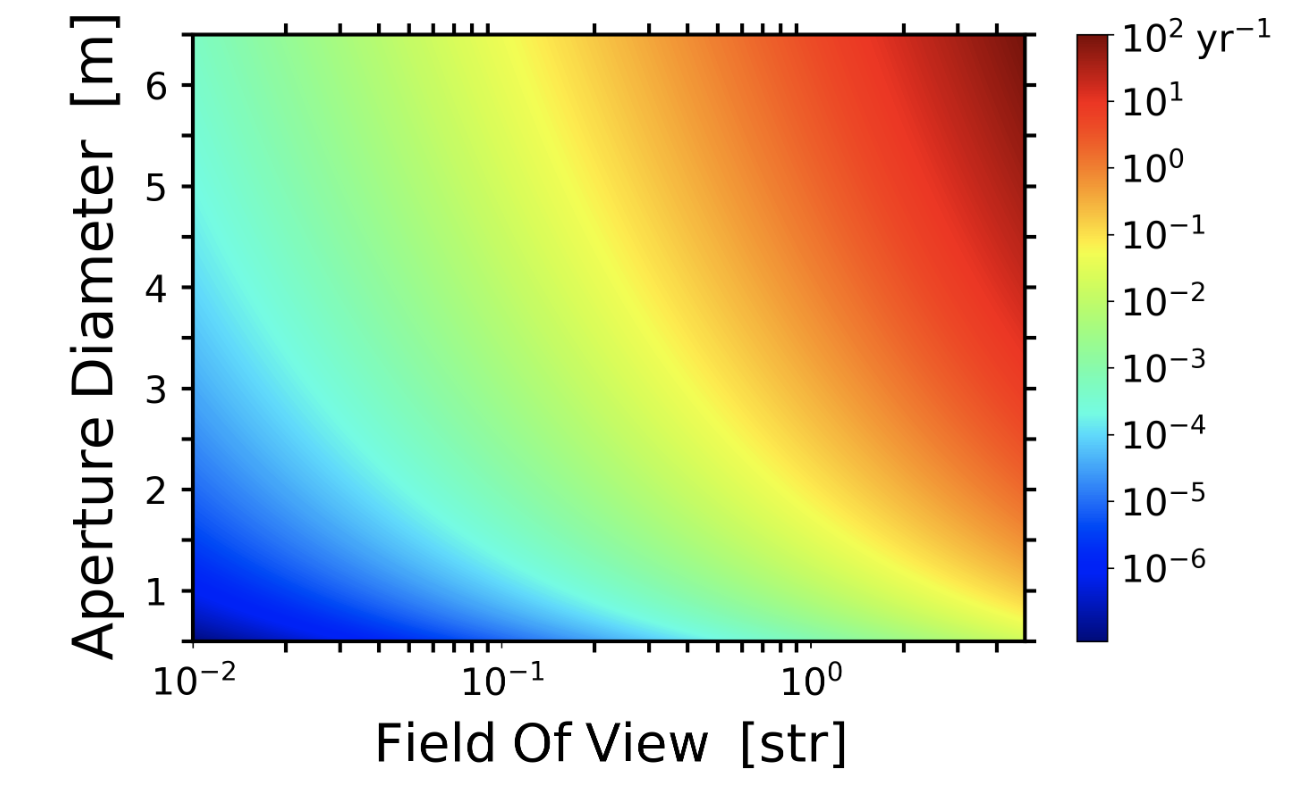}
    \caption{Expected interstellar impact rate as a function of telescope aperture diameter and field of view.}
    \label{fig:jet}
\end{figure}

Figure~\ref{fig:jet} shows the expected interstellar impact rate for a telescope in orbit 100 km above the lunar surface, calibrated based on CNEOS 2014-01-08 \citep{2019arXiv190407224S, 2019arXiv190603270S}. A telescope with an aperture diameter of size $D_a \gtrsim 2 \mathrm{\; m}$ and maximal FOV ($\sim 4.9 \mathrm{\; str}$) is expected to observe $\gtrsim 1$ interstellar meteoroid impact per year. For each meteoroid, measurements of the reflected sunlight and shadow, as well as the resulting optical flash and crater, would allow for the determination of velocity, mass, density, radiative efficiency, and composition. It is important to note that the interstellar number density calibration is based upon a single detection of CNEOS 2014-01-08, and therefore the appropriate Poisson statistics apply. The 95\% confidence bounds for an estimate of $\sim 1 \; \mathrm{yr^{-1}}$ are $0.03 - 5.57 \mathrm{\; yr^{-1}}$. Meteoroids with known velocities and directions, such as those associated with meteor showers \citep{2000ApJ...542L..65B, 2015MNRAS.454..344O, 2019MNRAS.484.5212A}, can help calibrate the system. Non-interstellar meteoroids outnumber interstellar ones by a factor of nearly $\sim 10^3$ in the centimeter-size scale \citep{2006mess.book.....L}, so a $D_a \sim 2 \mathrm{\; m}$ telescope that is capable of detecting a single interstellar meteoroid impact per year should detect at least one non-interstellar meteoroid per day. This provides the opportunity to study hypervelocity impacts on a daily basis, constraining the associated cratering and radiative processes \citep{2000ApJ...542L..65B, 2001ESASP.495..533D, 2010EM&P..107...55B, 2015MNRAS.454..344O, 2006Icar..184..319O, GOEL201554, 2015JGRE..120.1847S, 2019MNRAS.484.5212A}. In addition, follow-up studies of fresh craters can be performed by lunar rovers, revealing the physics of hypervelocity impacts as well as the composition of the meteoroid in more detail.

Given that NASA's Artemis program plans to  establish a sustainable base on the Moon\footnote{\url{https://www.nasa.gov/specials/artemis/}}, we also explored the detection signatures of interstellar dust on its regolith and rocks \cite{2020ApJ...895L..42S}. There is evidence that nearby supernovae have resulted in the $^{60}\mathrm{Fe}$ and other radionuclides detected in deep-ocean samples \citep{1999PhRvL..83...18K, 2004PhRvL..93q1103K, 2012PASA...29..109F, 2016Natur.532...69W}, the lunar surface \citep{2016PhRvL.116o1104F}, cosmic rays \citep{1974Sci...184.1079R, 2015PhRvL.115r1103K, 2018PhRvD..97f3011K}, and microfossils \citep{2016PNAS..113.9232L}. Supernovae eject can accelerate dust to sub-relativistic speeds $\lesssim 0.1 \mathrm{\; c}$ \citep{1949PhRv...76..583S, 1999APh....12...35B, 2003LNP...598.....W, 2015ApJ...806..255H}. Studying dust accelerated by supernovae could elucidate the history of nearby supernovae \citep{2004PhRvL..93q1103K, 2016ApJ...826L...3T, 2016Natur.532...69W, 2017ApJ...840..105M} and constrain theoretical models of supernovae \citep{2015MNRAS.446.2089W, 2017MNRAS.465.4044B, 2017MNRAS.465.3309D, 2019MNRAS.489.4465K}. 

We found that tracks resulting from $r \sim 0.01 - 0.07 \mathrm{\; \mu m}$ dust grains accelerated by supernovae to speeds of $r \sim 0.01 - 0.03 \mathrm{\; c}$ can be discovered in the lunar rocks. Studies of lunar rocks could shed light on the history of supernovae within the past $\sim 1 \mathrm{\; Myr}$ and within $\sim 250 \mathrm{\; pc}$, with the potential to reveal the timings, lumonsities, and directions of recent supernovae. The expected density of tracks is $\sim 10^{3} \; \mathrm{cm^{-2}}$ for depths of $\sim 0.1 \mathrm{\; mm}$ and widths of $\sim 0.01 \mathrm{\; \mu m}$ and $\sim 1 \; \mathrm{cm^{-2}}$ for depths of $\sim 0.7 \mathrm{\; mm}$ and widths of $\sim 0.07 \mathrm{\; \mu m}$. 
However, since lunar rocks may be covered by regolith material with typical grain sizes of $\sim 100 \mathrm{\; \mu m}$ \citep{1991lsug.book.....H}, impacts of the sub-relativistic grains considered here could catastrophically disrupt such grains instead of forming tracks within a rock, potentially reducing the actual density of tracks observed on lunar rocks.

The six Apollo missions brought to Earth 2200 lunar rocks\footnote{\url{https://curator.jsc.nasa.gov/lunar/}} which could be searched these tracks. While cosmic ray tracks have been discovered in lunar samples \citep{1974GeCoA..38.1625D, 1980asfr.symp..331C, 1981InEPS..90..359B}, these tracks would be differentiable by their widths which would be at least an order of magnitude larger. Extraterrestrial artifacts, such as microscopic probes akin to Breakthrough Starshot\footnote{\url{https://breakthroughinitiatives.org/initiative/3}}, could also form such tracks.

\subsection{Neutron star impacts}
\label{neu}

Finally, violent collisions between ISOs and neutron stars could give rise to radio transients. Fast Radio Bursts (FRBs) are observed to have millisecond duration at a frequency of $\sim 1$ GHz \citep{2019MNRAS.487..491K}. A multitude of hypotheses exist to explain FRBs, many involving neutron stars (NSs) \citep{2019PhR...821....1P}. \cite{2015ApJ...809...24G} proposed that FRBs may be powered by collisions between NSs and asteroids/comets, and \cite{2016ApJ...829...27D} studied the acceleration and radiation mechanisms of ultra-relativistic electrons in such collisions while proposing that repeating FRBs could be explained by NSs traveling through asteroid belts. The \cite{2015ApJ...809...24G} hypothesis lacks a clear source of asteroids to power non-repeating FRBs. Rotating Radio Transients (RRATs) have durations and frequencies similar to FRBs, but they originate in Milky Way galaxy and their luminosities are a billion times fainter \citep{2006Natur.439..817M}. In \cite{2019RNAAS...3..130S}, we explore the possibility that ISAs could power a subclass of RRATs which do not repeat over long timescales.

\begin{figure}[hptb]
  \centering
  \includegraphics[width=.7\linewidth]{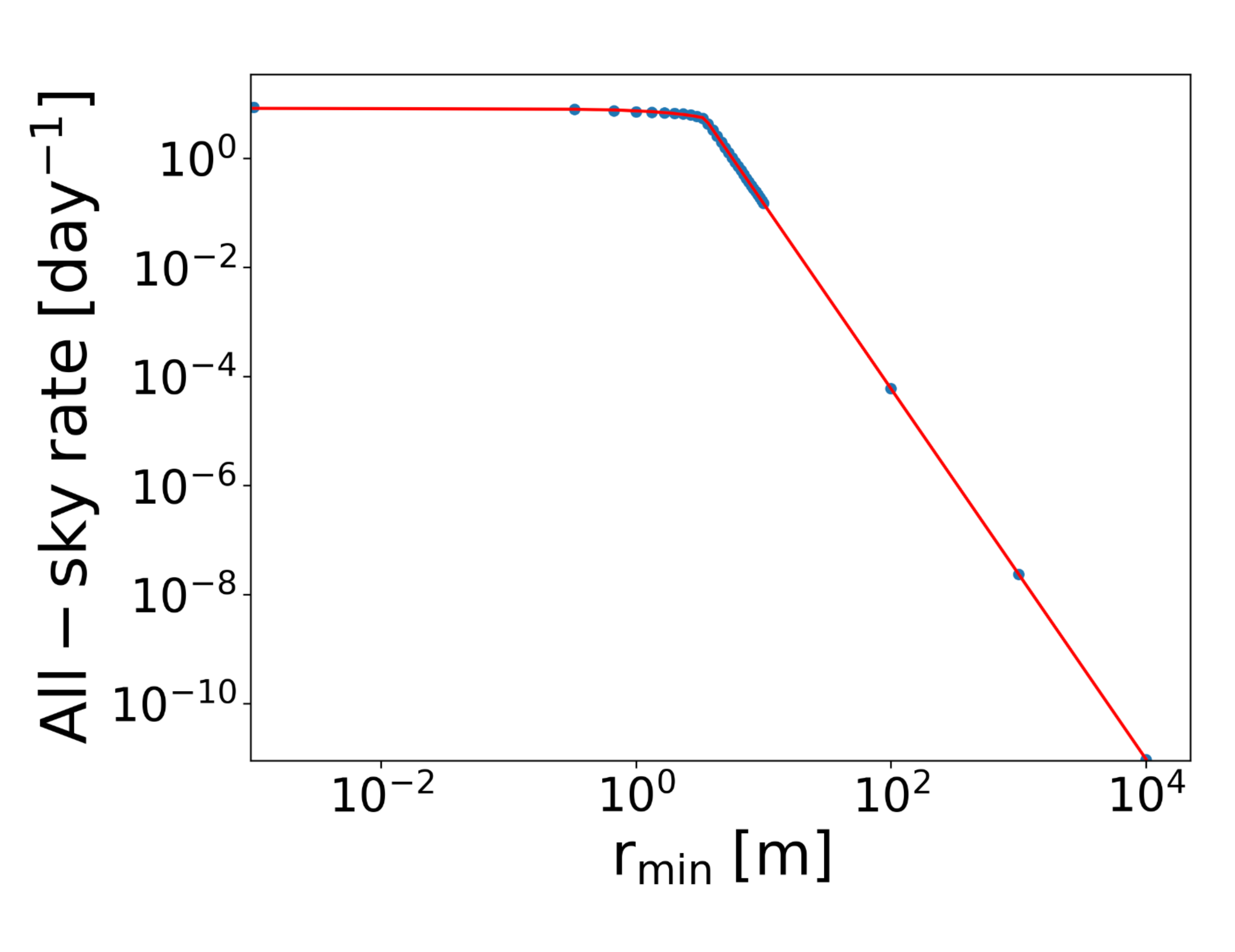}
    \caption{Rate of $\sim 1$ Jy radio flares at $\nu \sim$ 1 GHz from NS-ISA collisions, as a function of the minimum ISA radius, $r_{min}$.}
    \label{fig:3}
\end{figure}

We find that NS-ISA collisions could reliably power observable, non-repeating, millisecond-duration $\sim 1$ GHz radio flares in the Milky Way galaxy, the distribution of which is shown in Figure \ref{fig:3}. We would not expect to detect any X-ray emission from such events, given the expression for X-ray flux in \cite{2016ApJ...829...27D} for $r_{min} \gtrsim 1 \; \mathrm{m}$. We do not expect such events to constitute a significant fraction of FRBs due to the low abundance of sufficiently large asteroids to produce observable flares at cosmological distances. The abundance of single-pulse RRATs is still poorly constrained, so it is difficult to compare our estimated rate with the total estimated rate (Agarwal, McLaughlin \& Lorimer, private communication). NS-ISA collisions represent a new class of transients that could reveal the distributions and abundances of both NSs and ISAs, serving as an important calibration for both populations.

\section{Future observations of ISOs}
\label{fut}

One of the two branches of the Galileo Project is focused on obtaining better data on ISOs in an attempt to identify the nature of objects that do not resemble comets or asteroids, like `Oumuamua. \footnote{\url{https://projects.iq.harvard.edu/galileo}} The initiative will aim to develop software that will be applied to pipelines of archival and future surveys as well as the design of space missions that will investigate ISOs.

\begin{figure}[hptb]
  \centering
  \includegraphics[width=0.7\linewidth]{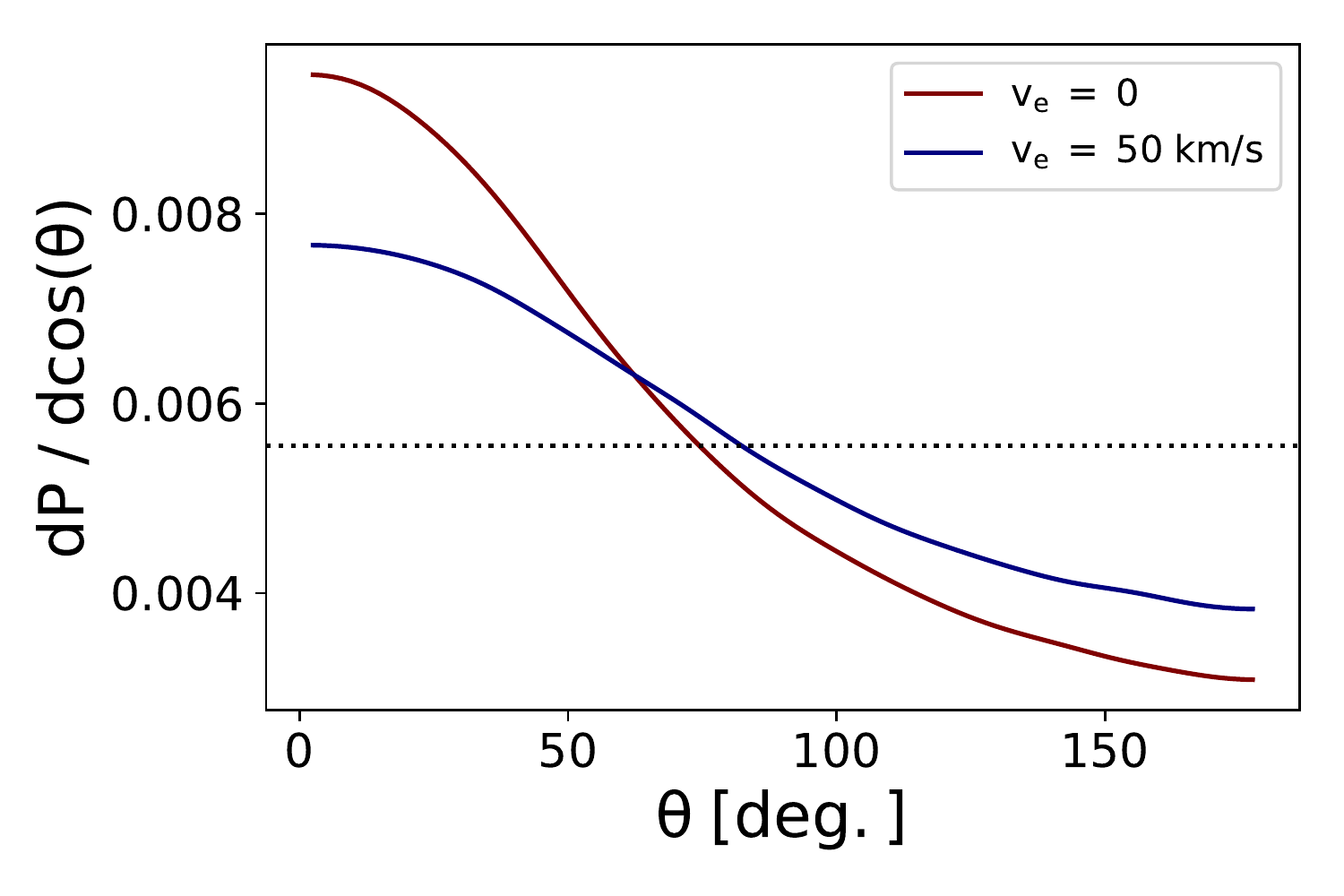}
    \caption{Normalized differential probability distributions, for ejection speeds of $0$ and $50 \; \mathrm{km \; s^{-1}}$, as a function of angular velocity vector displacement from the Solar antapex. If the Sun were stationary to the LSR, the distribution of $dP / d \cos{(\theta)}$ would be uniform, as indicated by the dotted line.
}
    \label{fig:eject}
\end{figure}

The LSST will greatly increase the cadence of ISO discovery in the solar system. If ISOs originate from stars, their velocity distribution combines the stellar and the ejection speed velocity distributions. The three-dimensional velocity dispersion ellipsoid of local stars relative to the local standard of rest (LSR) is, $(\sigma_U, \sigma_V, \sigma_W) = (33 \pm 4, 38 \pm 4, 23 \pm 2) \; \mathrm{km \; s^{-1}}$ \citep{2018MNRAS.474..854A}.\footnote{\label{lsr}Quoted errors represent 1-sigma uncertainties. Adopting alternative definitions of the LSR, such as only averaging over dwarf stars, may alter the results described in \cite{2020ApJ...903L..20S} by $< 20 \%$ \citep{2017RNAAS...1...21M}.} Since ejection speeds are comparable to pre-ejection orbital speeds, they reflect the region from which ISOs are likely produced in planetary systems. This information is essential for understanding how ISOs are produced. Because of the motion of the Sun relative to local stars, $(v_U^{\odot}, v_V^{\odot}, v_W^{\odot}) = (10 \pm 1, 11 \pm 2, 7 \pm 0.5) \; \mathrm{km \; s^{-1}}$ away from the LSR \citep{2010MNRAS.403.1829S, 2015ApJ...809..145T, 2016ARA&A..54..529B}, there should be an anisotropy in the distribution of ISOs. \cite{2020ApJ...903L..20S} demonstrates this prediction through numerical simulations, and furthermore, shows that a larger ejection speed of ISOs from their parent stars will lead to an increase in the predicted $v_{\infty}$ distribution and reduction in the predicted angular anisotropy. The difference between a low- and high-ejection-speed ISOs will be observable by LSST and the overall degree of angular anisotropy will be revealed best by ISOs with velocity vectors similar to that of the Solar antapex. Figure \ref{fig:eject} shows the normalized probability distributions, adjusted for the solid angle associated with $\theta$ for $0$ and $50 \mathrm{\; km \; s^{-1}}$ ejection speed cases. A systematically larger ejection speed leads to a flattened distribution of $d\mathrm{P} / d \cos{\mathrm{\theta}}$. With a sufficiently large number of ISOs, it should be possible to infer the characteristic value of the typical ejection speed. 

Interstellar asteroids, characterized by their rocky compositions, are expected to have high ejection speeds from their parent stars, and interstellar comets, characterized by their icy compositions, are expected to have low ejection speeds from their parent stars, since the two populations originate from the inner and outer regions, respectively, of planetary systems. The most natural expectation is that most ISOs originate from exo-Oort clouds since they represent the regions of exoplanetary systems with the lowest gravitational binding energies. As a result, these objects would be interstellar comets, with icy compositions, and with ejection speeds near zero. Interactions with turbulent source environments as well as the interstellar medium may provide additional constraints on ISO velocities \citep{2020ApJ...894...36L}. The complete velocity distribution of ISOs may contain multiple components that are distinguishable, with ejection speeds ranging from leaving outer Oort clouds around stars at low velocities ($v_e \sim 0$) up to escaping the tidal disruption orbital region near white dwarfs ($v_e \sim \mathrm{\; hundreds \; km \; s^{-1}}$). LSST will be capable of calibrating the relative mass fraction in each of these components. If extraterrestrial artifacts constitute a sub-population of ISOs \citep{2018ApJ...868L...1B}, they may also be distinguishable using the twin measures of velocity dispersion and angular anisotropy.

\begin{figure}
  \centering
  \includegraphics[width=0.7\linewidth]{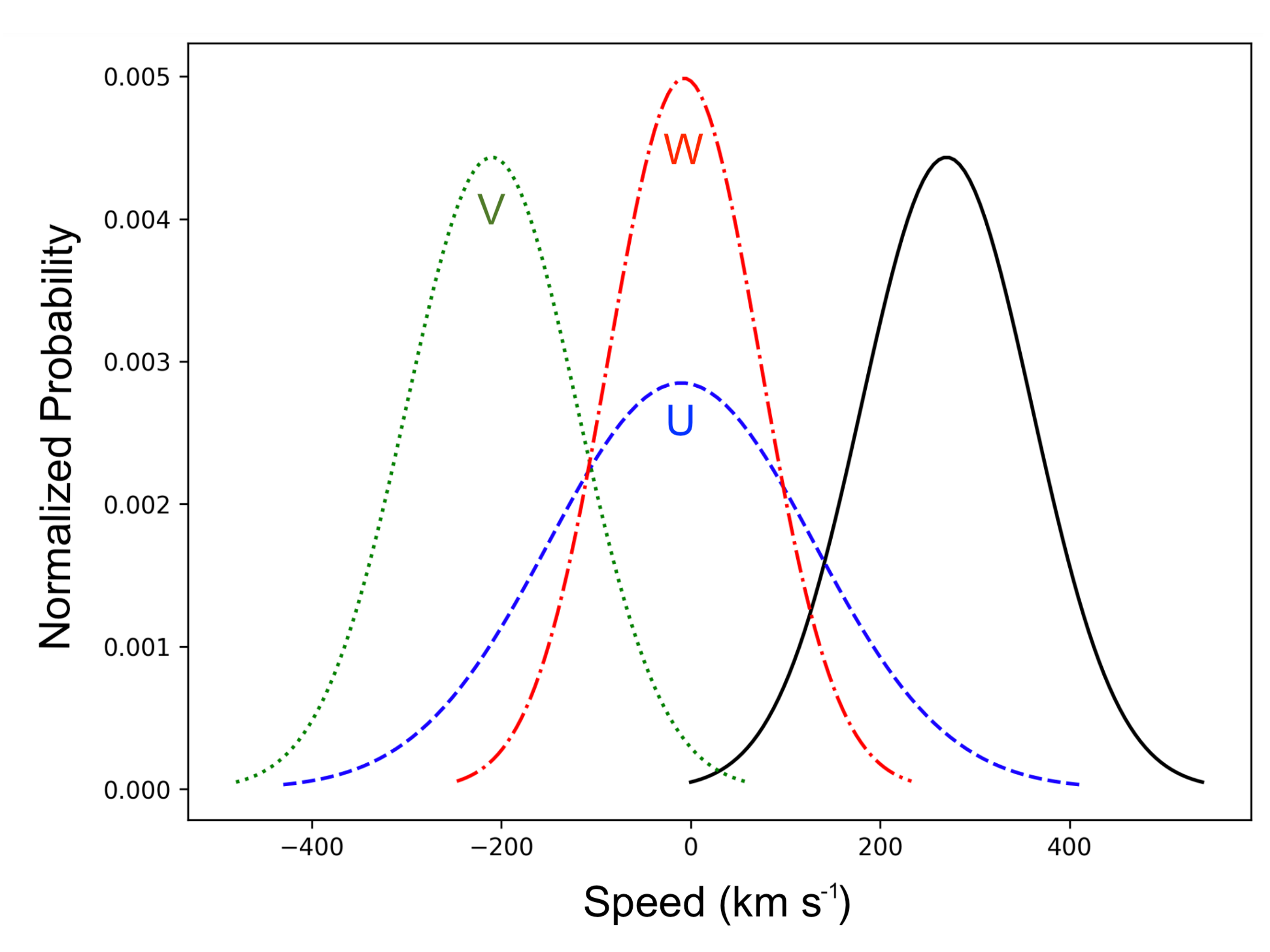}
    \caption{Distribution of heliocentric impact velocities of halo ISOs, with U, V, and W, components shown in blue, green, and red, respectively, and total speed displayed in black. Cutoffs are at $\pm 3 \sigma$, where $\sigma$ is the standard deviation.}
    \label{fig:1}
\end{figure}

In addition to the insights that can be drawn from large-number statistics, future observations will also reveal rare sub-classes of ISOs that carry unique insights. In \cite{2021NewA...8401545S}, we explore the theoretical population of ISOs originating from the stellar halo, which contains the oldest stars in the Milky Way galaxy and in the universe \citep{2008A&ARv..15..145H, 2007ApJ...660L.117F}. Studying the composition of halo star ejecta could help reveal the nature of primordial planetary system formation. We find the heliocentric impact velocity distribution for halo ISOs to have mean velocities $(\bar{v}_U^h, \bar{v}_V^h, \bar{v}_W^h) = (-10, -210, -7) \mathrm{\;km\;s^{-1}}$, with velocity dispersions $(\sigma_U^h, \sigma_V^h, \sigma_W^h) = (140, 90, 80) \mathrm{\;km\;s^{-1}}$, shown  in Figure~\ref{fig:1}. The mean heliocentric impact speed is $\sim 270 \mathrm{\;km\;s^{-1}}$, and the standard deviation is $\sim 90 \mathrm{\;km\;s^{-1}}$, making halo ISOs kinematically distinct from all other (Solar System or interstellar) meteors which have characteristic impact speeds that are an order-of-magnitude smaller.

Approximately 1 in every $10^3$ local stars, and therefore interstellar meteors, originate from the Milky Way halo \citep{2008A&ARv..15..145H}. On average, halo ISOs travel $\sim5$ times faster than disk ISOs. We therefore estimate the impact rates for halo ISOs to be a factor of $\sim 200$ lower than those for typical interstellar meteors of the same size \citep{2021NewA...8401545S}. All-sky camera systems such as AMOS can obtain spectra for typical meteors down to a size scale of $\sim 1$cm, and considering that halo interstellar meteor impact speeds are approximately ten times higher than typical meteors, we estimate that, with an adequate frame rate, a system with the sensitivity of AMOS could obtain spectra for halo ISOs down to a size scale of $\sim2$mm \citep{2015P&SS..118..102T}. We expect 2mm halo ISOs to impact the Earth at a rate of $10^4 \mathrm{\;yr^{-1}}$. A network of all-sky cameras covering all land on Earth, with the sensitivity of AMOS and an adequate frame rate to determine the orbits of $\sim 300 \mathrm{\;km\;s^{-1}}$ impactors, are expected to detect and take spectra for nearly $10^3$ halo ISO per year. We would not expect to obtain physical samples of halo ISOs, as meteors larger than 10cm are expected to impact the Earth once every few hundred years. Finally, the ratio between objects with disk versus halo origins will inform our understanding of primoridal planetary system formation.

The stellar halo is home to the oldest stars in the Milky Way and in the universe \citep{2008A&ARv..15..145H, 2007ApJ...660L.117F}. Studying the composition of their ejecta can therefore reveal the formation history of some of the first planets, as well as the enrichment history of the early universe. For instance, the observed ratio of icy to rocky planetesimals can help constrain models of planetary system formation, the chemical composition of rocky planetesimals could reveal unknown details about the gas-clearing and planet migration phases, and the ratios of volatiles in icy planetesimals  could inform the prospects for life in early planetary systems. Compositional information for halo ISOs could also help constrain the origins of exo-Oort clouds, as well as chemical diversity among early planetary systems. Furthermore, analysis of halo ISOs could test the possibility that carbon-enhanced metal-poor poor stars could create carbon planets \citep{2016MNRAS.460.2482M}.

Given the science described in this review, the future of ISO research is an exciting new frontier in astronomy. The ISO branch of the Galileo Project will learn more about the nature of future ISOs like `Oumuamua by developing new searches and follow-up observations for these unusual objects.

\section*{Acknowledgments}

This research was supported in part by a grant from the Breakthrough Prize Foundation as well as funds from the Galileo Project at Harvard University.

\bibliographystyle{abbrvnat}
\bibliography{bib}

\end{document}